\documentclass[12pt]{iopart}
\usepackage{iopams}  
\usepackage{lscape}
\usepackage{graphicx}
\usepackage{threeparttable}
\usepackage{multirow}
\usepackage{color}
\newcommand{\mnras}{MNRAS}
\newcommand{\aap}{Astronomy and Astrophysics}

\newcommand{\apj}{ApJ}

\newcommand{\prd}{Phys. Rev. D}

\newcommand{\hatwo}{\hat{\alpha}_2}

\newcommand{\hxi}{\hat{\xi}}
\newcommand{\psrb}{PSR~B1937+21}
\newcommand{\psrj}{PSR~J1744$-$1134}
\newcommand{\psrs}{PSRs~B1937+21 and J1744$-$1134}
\begin{document}

\title[New limits on LPI violation of gravity]{New limits on
the violation of local position invariance of gravity}

\author{Lijing Shao$^{1,2}$ and Norbert Wex$^{1}$}

\address{$^1$\,Max-Planck-Institut f\"{u}r Radioastronomie, Auf dem
  H\"{u}gel 69, D-53121 Bonn, Germany} 
\address{$^2$\,School of
  Physics, Peking University, Beijing 100871, China}
  
\eads{\mailto{lshao@pku.edu.cn (LS)},
  \mailto{wex@mpifr-bonn.mpg.de (NW)}}

\begin{abstract}

Within the parameterized post-Newtonian (PPN) formalism, there could
be an anisotropy of local gravity induced by an external matter
distribution, even for a fully conservative metric theory of
gravity. It reflects the breakdown of the local position invariance of
gravity and, within the PPN formalism, is characterized by the
Whitehead parameter $\xi$. We present three different kinds of
observation, from the Solar system and radio pulsars, to constrain
it. The most stringent limit comes from recent results on the
extremely stable pulse profiles of solitary millisecond pulsars, that
gives $|\hxi| < 3.9 \times 10^{-9}$ (95\% CL), where the hat denotes
the strong-field generalization of $\xi$. This limit is six orders of
magnitude more constraining than the current best limit from
superconducting gravimeter experiments. It can be converted into an
upper limit of $\sim 4 \times 10^{-16}$ on the spatial anisotropy of
the gravitational constant.

\end{abstract}

%Uncomment for PACS numbers title message
% Keywords required only for MST, PB, PMB, PM, JOA, JOB? 
%\vspace{2pc}
%\noindent{\it Keywords}: gravitation, local position invariance,
%pulsars

% Uncomment for Submitted to journal title message
%\submitto{\JPA}
% Comment out if separate title page not required
\maketitle
%%%%%%%%%%%%%%%%%%%%%%%%%%%%%%%%%%%%%%%%%%%%%%%%%%%%%%%%%%%%%%%%%%%%%%

%=====================================================================
\section{Introduction}
\label{sec:intro}
%=====================================================================

Since the 1960s, advances in technologies are continuously providing a
series of formidable tests of gravity theories from on-ground
laboratories, the Solar system, various pulsar systems, and also
cosmology~\cite{wil93,wil06}. Up to now, Einstein's general relativity
(GR) passed all experimental tests with flying colors. However,
questions related to the nature of dark matter and dark energy, and
irreconcilable conflicts between GR and the standard model of particle
physics, are strong motivations to study alternative theories of
gravity.  In addition, gravity as a fundamental interaction of nature
deserves most stringent tests from various aspects.

For tests of gravity theories, one of the most popular frameworks is
the {\it parameterized post-Newtonian (PPN) formalism}, proposed by
Nordtvedt and Will~\cite{nor68b,wil71,wn72,wil93}. In the standard PPN
gauge, the framework contains ten dimensionless PPN parameters in the
metric components as coefficients of various potential forms. These
parameters take different values in different gravity theories. Hence,
experimental constraints on these parameters can be directly used to
test specific gravity theories~\cite{nw72,wil93,wil06}.

In this paper, we concentrate on one of the ten PPN parameters which
characterizes a possible Galaxy-induced anisotropy in the
gravitational interaction of localized systems. Such an anisotropy is
described by the Whitehead parameter $\xi$ in the weak-field
slow-motion limit~\cite{wil73}.  We use $\hxi$ to explicitly denote
its strong-field generalization. Besides Whitehead's gravity
theory~\cite{whi22}, $\xi$ is relevant for a class of theories called
``quasilinear'' theories of gravity~\cite{wil73}. In GR, the
gravitational interaction is local position invariant with $\xi = 0$,
while in Whitehead's gravity, local position invariance (LPI) is
violated and $\xi = 1$~\cite{wil73,gw08}.

An anisotropy of gravitational interaction, induced by the
gravitational field of the Galaxy, would lead to anomalous Earth tides
at specific frequencies with characteristic phase
relations~\cite{wil73,wg76}. The $\xi$-induced Earth tides are caused
by a change in the local gravitational attraction on the Earth surface
due to the rotation of the Earth with frequencies associated with the
sidereal day.  By using constraints on $\xi$ from superconducting
gravimeter, Will gave the first disproof of Whitehead's parameter-free
gravity theory~\cite{wil73} (see~\cite{gw08} for multiple recent
disproofs). Later Warburton and Goodkind presented an update on the
limit of $\xi$ by using new gravimeter data~\cite{wg76}, where they
were able to constrain $|\xi|$ to the order of $10^{-3}$. The
uncertainties concerning geophysical perturbations and the imperfect
knowledge of the Earth structure limit the precision. Uncertainties
include the elastic responses of the Earth, the effects of ocean
tides, the effects of atmospheric tides from barometric pressure
variation, and the resonances in the liquid core of the
Earth~\cite{wg76} (see~\cite{goo99,shi08} for recent reviews on
superconducting gravimeters).

Limits from Earth tides are based on periodic terms proportional to
$\xi$, while secular effects in other astrophysical laboratories can
be more constraining. Nordtvedt used the close alignment of the Sun's
spin with the invariable plane of the Solar system to constrain the
PPN parameter $\alpha_2$, associated with the local Lorentz invariance
of gravity down to ${\cal O}(10^{-7})$~\cite{nor87}.  In the same
publication Nordtvedt pointed out that such a limit is also possible
for $\xi$, as the two terms in the Lagrangian have the same form.
However, to our knowledge, no detailed calculations have been
published yet. In section~\ref{sec:solar} we follow Nordtvedt's
suggestion and achieve a limit of ${\cal O}(10^{-6})$.

A non-vanishing (strong-field) $\hxi$ would lead to characteristic
secular effects in the dynamics of the rotation and orbital motion of
radio pulsars. We have presented the methodologies in details to
constrain the (strong-field) $\hatwo$ from binary pulsar
timing~\cite{sw12} and solitary pulsar profile analysis~\cite{sck+13}
respectively. By the virtue of the similarity between $\hatwo$- and
$\hxi$-related effects, in section~\ref{sec:bnry:psr} we extend the
analysis in~\cite{sw12,sck+13} to the case of LPI of gravity. From
timing results of PSRs~J1012+5307~\cite{lwj+09} and
J1738+0333~\cite{fwe+12}, a limit of $|\hxi| < 3.1 \times 10^{-4}$
(95\% CL) is achieved for neutron star (NS) white dwarf (WD) systems
\cite{swk12}.  As shown in this paper, from the analysis on the pulse
profile stability of \psrs{}, a limit of $|\hxi| < 3.9 \times 10^{-9}$
(95\% CL) is obtained, utilizing the rotational properties of solitary
millisecond pulsars. This limit is six orders of magnitude better than
the (weak-field) limit from gravimeter.

The paper is organized as follows. In the next section, the
theoretical framework for tests of LPI of gravity is briefly
summarized. In section~\ref{sec:solar}, a limit on $\xi$ from the
Solar system is obtained. Then we give limits on $\hxi$ from binary
pulsars and solitary pulsars in section~\ref{sec:bnry:psr}.  In the
last section, we discuss issues related to strong-field modifications
and conversions from our limits to limits on the anisotropy in the
gravitational constant.  Comparisons between our tests with other
achievable tests from gravimeter and lunar laser ranging (LLR)
experiments are also given.

%=====================================================================
\section{Theoretical framework}
\label{sec:th}
%=====================================================================

In the PPN formalism, PPN parameters are introduced as dimensionless
coefficients in the metric in front of various potential
forms~\cite{wn72,wil93,wil06}. In the standard post-Newtonian gauge,
$\xi$ appears in the metric components $g_{00}$ and
$g_{0i}$~\cite{wn72,wil93,wil06}. However, in most cases, it is
relevant only in linear combinations with other PPN parameters like
$\beta$, $\gamma$ (see~\cite{wil93,wil06} for formalism and
details). Due to the limited precision in constraining these PPN
parameters (see table~4 in~\cite{wil06} for current constraints on PPN
parameters), it is not easy to get an independent stringent limit for
$\xi$. For example, based on the Nordtvedt parameter (see (43)
in~\cite{gw08}),
%% ===================================================================
\begin{equation}
  \eta = 4\beta - \gamma - 3 - \frac{10}{3}\xi - \alpha_1 +
  \frac{2}{3} \alpha_2 - \frac{2}{3}\zeta_1 - \frac{1}{3} \zeta_2 \,,
\end{equation}
%% ===================================================================
one can only constrain $\xi$ to the order of ${\cal O}(10^{-3})$ at
most.  Nevertheless, in the metric component $g_{00}$, $-2\xi$ alone
appears as the coefficient of the Whitehead potential~\cite{wil73},
%% ===================================================================
\begin{equation}\label{eq:Phiw}
  \Phi_W ({\bf x}) \equiv \frac{G^2}{c^2} \int\hspace{-0.3cm}\int
  \rho({\bf x}') \rho({\bf x}'') \left(\frac{{\bf x}-{\bf x}'}{|{\bf
      x}-{\bf x}'|^3}\right) \cdot \left( \frac{{\bf x}'-{\bf
      x}''}{|{\bf x}-{\bf x}''|} - \frac{{\bf x}-{\bf x}''}{|{\bf
      x}'-{\bf x}''|}\right) {\rm d}^3 {\bf x}' {\rm d}^3 {\bf x}''
  \,,
\end{equation}
%% ===================================================================
where $\rho({\bf x})$ is the matter density, $G$ and $c$ are the
gravitational constant and the speed of light respectively. This fact
provides the possibility to constrain the PPN parameter $\xi$
directly.

Correspondingly, in the PPN $n$-body Lagrangian, we have a
$\xi$-related term for three-body interactions (see e.g. (6.80)
in~\cite{wil93}),
%% ===================================================================
\begin{equation}\label{eq:lagrangian3}
  L_\xi = - \frac{\xi}{2} \, \frac{G^2}{c^2} \sum_{i,j} \frac{m_i
    m_j}{r_{ij}^3} \, {\bf r}_{ij} \cdot \left[\sum_{k} m_k \left(
    \frac{{\bf r}_{jk}}{r_{ik}} - \frac{{\bf r}_{ik}}{r_{jk}} \right)
    \right] \,,
\end{equation}
%% ===================================================================
where the summation excludes terms that make any denominators vanish.
For our purposes below, we consider the third body being our Galaxy,
and only consider a system ${\cal S}$ (the Solar system or a pulsar
binary system or a solitary pulsar) of typical size much less than its
distance to the Galactic center $R_{\rm G}$.  Hence the Lagrangian
(\ref{eq:lagrangian3}) reduces to (dropping a constant factor that
rescales $G$)
%% ===================================================================
\begin{equation}\label{eq:lagrangianG}
  L_\xi = \frac{\xi}{2} \, \frac{U_{\rm G}}{ c^2} \sum_{i,j}
  \frac{Gm_im_j}{r_{ij}^3} ({\bf r}_{ij} \cdot {\bf n}_{\rm G})^2 \,,
\end{equation}
%% ===================================================================
where $U_{\rm G}$ is the Galactic potential at the position of the
system ${\cal S}$ (associated with the mass inside $R_{\rm G}$), and
${\bf n}_{\rm G} \equiv {\bf R}_{\rm G}/R_{\rm G}$ is a unit vector
pointing from ${\cal S}$ to the Galactic center. In our calculations
below we will use $U_{\rm G} \sim v_{\rm G}^2$, where $v_{\rm G}$ is
the rotational velocity of the Galaxy at ${\cal S}$.
Equation~(\ref{eq:lagrangianG}) is exact, only if the external mass is
concentrated at the Galactic center, otherwise a correcting factor has
to be applied, which depends on the model for the mass distribution in
our Galaxy~\cite{mn96}. At the end of section~\ref{sec:sum}, we show
that this factor is close to two, as already estimated in \cite{gw08}.

From Lagrangian (\ref{eq:lagrangianG}), a binary system of mass $m_1$
and $m_2$ gets an extra acceleration for the relative movement (see
(8.73) in~\cite{wil93} with different sign conventions),
%% ===================================================================
\begin{equation}
  {\bf a}_\xi = \xi \frac{U_{\rm G}}{c^2} \frac{G(m_1+m_2)}{r^2}
  \left[ 2({\bf
      n}_{\rm G} \cdot {\bf n}){\bf n}_{\rm G} -3{\bf n}\left({\bf
      n}_{\rm G} \cdot {\bf n}\right)^2 \right],
\end{equation}
%% -----------------------------------------------------------------
where ${\bf r} \equiv {\bf r}_1 - {\bf r}_2$ and ${\bf n} \equiv {\bf
  r}/r$. Because of the analogy between the extra acceleration caused
by the PPN parameter $\alpha_2$ (see (8.73) in~\cite{wil93}), the
Lagrangian (\ref{eq:lagrangianG}) results in similar equations of
motion with replacements,
%%--------------------------------------------------------------------
\begin{equation}\label{eq:replace}
  {\bf w} \to {\bf v}_{\rm G} \quad \mbox{and} \quad \alpha_2 \to -2\xi \,,
\end{equation}
%%--------------------------------------------------------------------
where ${\bf v}_{\rm G} \equiv v_{\rm G} {\bf n}_{\rm G}$ is an
effective velocity~\cite{swk12}. With replacements (\ref{eq:replace}),
the influence of $\xi$ for an eccentric orbit of a binary system can
be read out readily from (17--19) in~\cite{sw12}.  As for the
$\alpha_2$ test, in the limit of small eccentricity, $\xi$ induces a
precession of the orbital angular momentum around the direction ${\bf
  n}_{\rm G}$ with an angular frequency~\cite{sw12},
%%--------------------------------------------------------------------
\begin{equation}\label{eq:bnryprec}
  \Omega^{\rm prec} = \xi \left(\frac{2\pi}{P_b}\right) \left(
  \frac{v_{\rm G}}{c}\right)^2 \cos\psi \,,
\end{equation}
%%--------------------------------------------------------------------
where $P_b$ is the orbital period, and $\psi$ is the angle between
${\bf n}_{\rm G}$ and the orbital angular momentum. This precession
would introduce observable effects in binary pulsar timing experiments
(see section~\ref{sec:bnry}).

Similar to the case of a binary system, for an isolated, rotating
massive body with internal equilibrium, Nordtvedt showed
in~\cite{nor87} that $\xi$ would induce a precession of the
spin around ${\bf n}_{\rm G}$ with an angular frequency (note,
in~\cite{nor87} $\xi^{\rm Nordtvedt} = - \frac{1}{2} \xi$),
%%--------------------------------------------------------------------
\begin{equation}\label{eq:solitaryprec}
  \Omega^{\rm prec} = \xi \left( \frac{2\pi}{P}\right) \left(
  \frac{v_{\rm G}}{c}\right)^2 \cos\psi \,,
\end{equation}
%% -----------------------------------------------------------------
where now $\psi$ stands for the angle between the spin of the body and
${\bf n}_{\rm G}$. This precession can be constrained by observables
in the Solar system and solitary millisecond pulsars (see
section~\ref{sec:solar} and section~\ref{sec:psr} respectively).

%=====================================================================
\section{A weak-field limit from the Solar spin}
\label{sec:solar}
%=====================================================================

%---------------------------------------------------------------------
\begin{figure}
  \begin{center}
    \includegraphics[width=10cm]{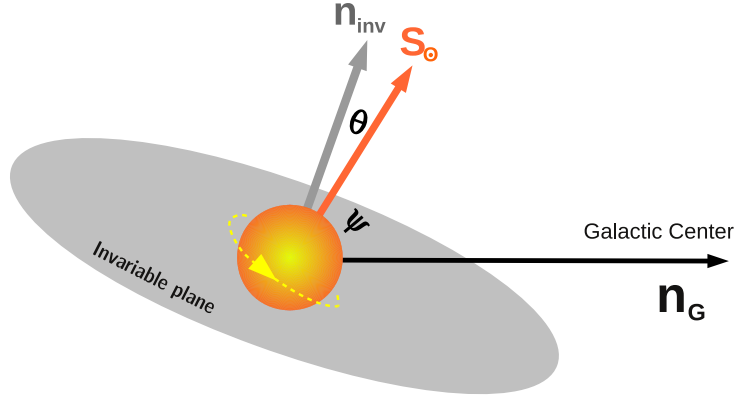}
  \end{center}
\caption{\label{fig:solar_geometry} Local position invariance
  violation causes a precession of the Solar angular momentum ${\bf
    S}_\odot$ around the direction of the local Galactic acceleration
  ${\bf n}_{\rm G}$, which causes characteristic changes in the angle
  $\theta$ between ${\bf S}_\odot$ and the norm of the invariable
  plane ${\bf n}_{\rm inv}$. Due to the movement of the Solar system
  in the Galaxy, ${\bf n}_{\rm G}$ is changing periodically with a
  period of $\sim250$\,Myr.}
\end{figure}
%---------------------------------------------------------------------

At the birth of the Solar system $\sim4.6$ billion years ago, the
angle $\theta$ between the Sun's spin ${\bf S}_\odot$ and the total
angular momentum of the Solar system (its direction is represented by
the norm of the invariable plane ${\bf n}_{\rm inv}$) were very likely
closely aligned, as suggested by our understanding of the formation of
planetary systems. After the birth, the Newtonian torque on the Sun
produced by the tidal fields of planets is negligibly weak (see
(\ref{eq:j2})).  Due to today's observation of $\theta \sim 6^\circ$,
Nordtvedt suggested to constrain $\xi$ to a high precision through
constraining (\ref{eq:solitaryprec})~\cite{nor87}. Based on his
$\alpha_2$ test and an order-of-magnitude estimation, he already
concluded $\xi \lesssim 10^{-7}$. Here we slightly improve his method
and present detailed calculations to constrain $\xi$ from the Solar
spin.

For directions of ${\bf S}_\odot$ and ${\bf n}_{\rm inv}$, we take the
International Celestial Reference Frame (ICRF) equatorial coordinates
at epoch J2000.0 from recent reports of the IAU/IAG Working Group on
Cartographic Coordinates and Rotational Elements~\cite{sah+07,ahb+11}.
The direction of ${\bf S}_\odot$ is $(\alpha_0,\delta_0)_\odot =
(286^\circ\!\!.13, 63^\circ\!\!.87)$ in the Celestial coordinates or
$(l,b)_\odot = (94^\circ\!\!.45, 22^\circ\!\!.77)$ in the Galactic
coordinates.  The coordinates of ${\bf n}_{\rm inv}$ are
$(\alpha_0,\delta_0)_{\rm inv} = (273^\circ\!\!.85,66^\circ\!\!.99)$
or $(l,b)_{\rm inv} = (96^\circ\!\!.92, 28^\circ\!\!.31)$.  The
difference between these two directions is
%---------------------------------------------------------------------
\begin{equation}\label{eq:misangle}
  \theta|_{t=0} = 5^\circ\!\!.97 \,,
\end{equation}
%---------------------------------------------------------------------
where $t=0$ denotes the current epoch. 

%---------------------------------------------------------------------
\begin{figure}
  \begin{center}
    \includegraphics[width=12cm]{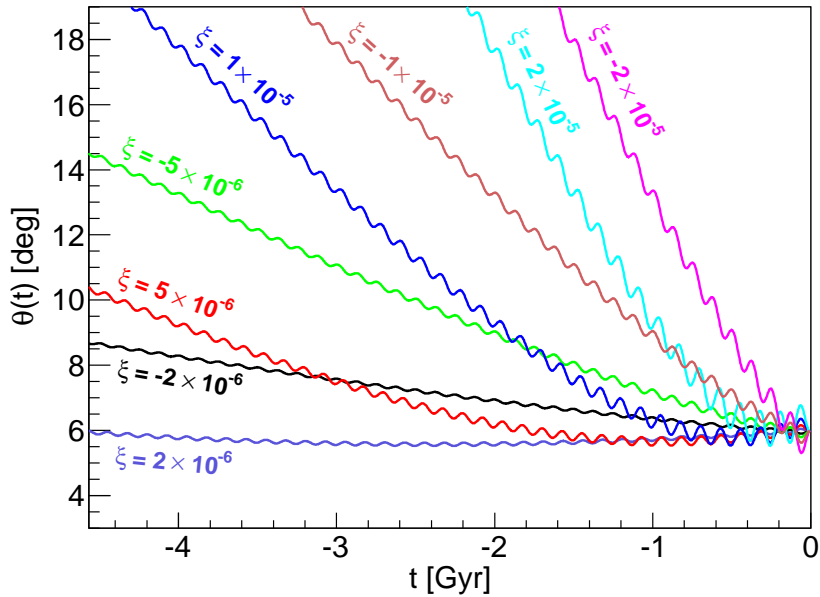}
  \end{center}
\caption{\label{fig:xis_solar} Evolutions of the misalignment angle
  $\theta(t)$ backward in time with different $\xi$ vaules, which have
  taken both (\ref{eq:solitaryprec}) and (\ref{eq:j2}) into account.}
\end{figure}
%---------------------------------------------------------------------

Assuming that the Sun's spin was closely aligned with ${\bf n}_{\rm
  inv}$ right after the formation of the Solar system, 4.6\,Gyr in the
past, one can convert (\ref{eq:misangle}) into a limit for $\xi$. For
this, one has to account for the Solar movement around the Galactic
center ($\sim 20$ circles in 4.6\,Gyr) when using
(\ref{eq:solitaryprec}) to properly integrate back in time for a given
$\xi$.  We show evolutions of the misalignment angle $\theta(t)$ in
figure~\ref{fig:xis_solar} for different $\xi$ vaules. In calculations
in figure~\ref{fig:xis_solar}, besides the contribution
(\ref{eq:solitaryprec}), we also include the precession produced by
the Newtonian quadrupole coupling with an angular frequency,
%---------------------------------------------------------------------
\begin{equation}\label{eq:j2}
  \Omega^{\rm prec}_{J_2} = \frac{3}{2} J_2 \frac{G M_\odot
    R_\odot^2}{ |{\bf S}_{\odot}|} \sum_i \frac{m_i}{r_i^3} \,,
\end{equation}
%---------------------------------------------------------------------
where $M_\odot$ and $R_\odot$ are the Solar mass and the Solar radius,
$m_i$ and $r_i$ are the mass and the orbital size of body $i$ in the
Solar system, and $J_2 = (2.40 \pm 0.25) \times
10^{-7}$~\cite{flk+11}. The main contributions in (\ref{eq:j2}) are
coming from Jupiter, Venus and Earth. The coupling is very weak, and
(\ref{eq:j2}) has a precession period $\sim 9\times10^{11}$\,yr, hence
it precesses $\sim 2^\circ$ in 4.6\,Gyr (notice a factor of two
discrepancy with (15) in~\cite{nor87} mainly due to the use of a
modern $J_2$ value). Such a precession hardly modifies the
evolution of $\theta(t)$; besides, the precession (\ref{eq:j2}) is
around ${\bf n}_{\rm inv}$ which by itself does not change $\theta$.

\begin{figure}
  \begin{center}
    \includegraphics[width=10cm]{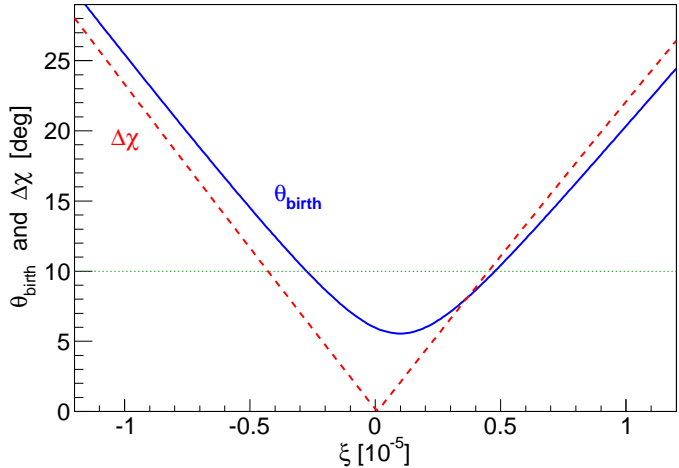}
  \end{center}
  \caption{\label{fig:solar_birth} The initial misalignment angle
    $\theta_{\rm birth}$ and the angle difference $\Delta\chi$ between
    current ${\bf S}_\odot$ and ${\bf S}_\odot$ at birth as functions
    of $\xi$.  They are obtained from evolving ${\bf S}_\odot$
    according to (\ref{eq:solitaryprec}) and (\ref{eq:j2}) back in
    time to the epoch $t=-4.6$\,Gyr.}
\end{figure}

In figure~\ref{fig:solar_birth} we plot the initial misalignment angle
at the birth of the Solar system and the angle $\Delta\chi$ swept out
by ${\bf S}_\odot$ during the past 4.6\,Gyr as functions of
$\xi$. From figure~\ref{fig:solar_birth} it is obvious that any $\xi$
significantly outside the range
%---------------------------------------------------------------------
\begin{equation}\label{eq:xi_solar}
  |\xi| \lesssim  5 \times 10^{-6} 
\end{equation}
%---------------------------------------------------------------------
would contradict the assumption that the Sun was formed spinning in a
close alignment with the planetary orbits (say, $\theta_{\rm birth}
\gtrsim 10^\circ$).  Limit (\ref{eq:xi_solar}) is three orders of
magnitude better than that from superconducting
gravimeter~\cite{wg76}.
%---------------------------------------------------------------------
%---------------------------------------------------------------------

%=====================================================================
\section{Limits from radio millisecond pulsars}
\label{sec:bnry:psr}
%=====================================================================

%=====================================================================
\subsection{A limit from binary pulsars}
\label{sec:bnry}
%=====================================================================

According to (\ref{eq:bnryprec}), the orbital angular momentum of a
binary system with a small eccentricity undergoes a $\xi$-induced
precession around ${\bf n}_{\rm G}$ (here ${\bf n}_{\rm G}$ is the
direction of the Galactic acceleration at the location of the
binary). As mentioned in~\cite{swk12}, this precession is analogous to
the precession induced by the PPN parameter $\alpha_2$~\cite{sw12}
with replacements (\ref{eq:replace}). Hence the same analysis done for
the $\hatwo$ test in~\cite{sw12} applies to the $\hxi$ test in binary
pulsars.

Using the Galactic potential model in~\cite{pac90} with the distance
of the Solar system to the Galactic center $\sim 8$\,kpc, Shao~\etal
\cite{swk12} performed $10^7$ Monte Carlo simulations to account for
measurement uncertainties and the unknown longitude of ascending node
(for details, see section 3 of~\cite{sw12}). From a combination of
PSRs~J1012+5307 and J1738+0333, they got a probabilistic limit (see
figure 1 in~\cite{swk12} for probability densities from separated
binary pulsars and their combination),
\begin{equation}\label{eq:xi_bnry}
  |\hxi| < 3.1 \times 10^{-4} \,, \quad \mbox{(95\% CL)}\,.
\end{equation}
%% -----------------------------------------------------------------
It is two orders of magnitude weaker than the limit
(\ref{eq:xi_solar}) from the Solar spin, but it represents a
constraint involving a strongly self-gravitating body, namely, NS-WD
binary systems (see section~\ref{sec:sum}).

%=====================================================================
\subsection{A limit from solitary pulsars}
\label{sec:psr}
%=====================================================================

Similar to the precession of the Solar spin, the spin of a solitary
pulsar would undergo a $\hxi$-induced precession around ${\bf n}_{\rm
  G}$ with an angular frequency (\ref{eq:solitaryprec}). Such a
precession would change our line-of-sight cut on the pulsar emission
beam, hence change the pulse profile characteristics over time, see
figure~1 in~\cite{sck+13} for illustrations.

Recently, to test the local Lorentz invariance of gravity, Shao
\etal~\cite{sck+13} analyzed a large number of pulse profiles from
\psrs{}, obtained at the 100-m Effelsberg radio telescope with the
same backend, spanning about $\sim15$ years. From various aspects, the
pulse profiles are very stable, and no change in the profiles is found
(see figures 2--7 in~\cite{sck+13} for stabilities of pulse profiles).
These results can equally well be used for a test of LPI of gravity.

\begin{figure}
  \begin{center}
    \includegraphics[width=10cm]{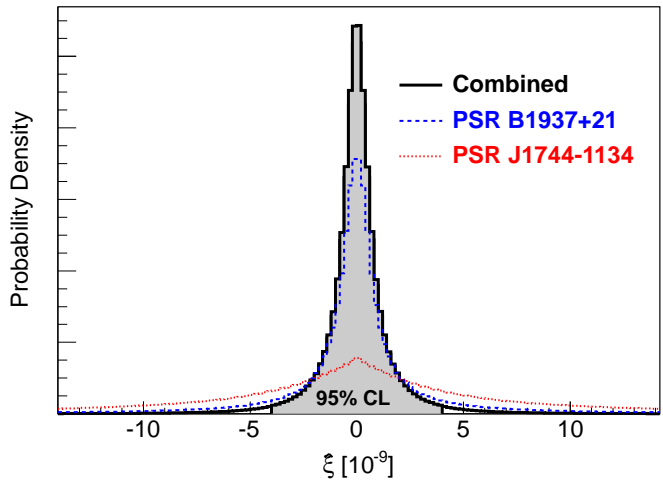}
  \end{center}
\caption{\label{fig:xi_psr} Probability density functions of the
  strong-field PPN parameter $\hxi$ from \psrb{} (blue dashed
  histogram), \psrj{} (red dotted histogram), and their combination
  (black solid histogram). All probability density functions are
  normalized.}
\end{figure}

By using a simple cone emission model of pulsars~\cite{lk05}, one can
quantitatively relate a change in the orientation of the pulsar spin
with that in the width of the pulse profile (see (10)
in~\cite{sck+13}). By using the limits on the change of pulse widths
in table 1 of~\cite{sck+13}, we set up $10^7$ Monte Carlo simulations
to get probability densities of $\hxi$ from \psrs{}. In simulations we
use the Galactic potential model in~\cite{pac90} and all other
parameters are the same as in~\cite{sck+13} with replacements
(\ref{eq:replace}). The results are shown in figure~\ref{fig:xi_psr}
for \psrs{} and their combination. For the individual limits one finds
%%====================================================================
\begin{eqnarray}
\mbox{\psrb{}:} && \quad 
   |\hxi| < 2.2 \times 10^{-8} \,, \quad\quad \mbox{(95\% CL)} \,, \\
\mbox{\psrj{}:} && \quad 
   |\hxi| < 1.2 \times 10^{-7} \,, \quad\quad \mbox{(95\% CL)} \,.
\end{eqnarray}
%%====================================================================
They are already significantly better than the limit
(\ref{eq:xi_solar}) obtained from the Solar spin. Like in
\cite{sck+13}, the analysis for \psrb{} is based on the
main-pulse. Also here, one could use the interpulse to constrain a
precession of \psrb{}, which again leads to a similar, even slightly
more constraining limit. As in \cite{sck+13}, we will stay with the
more conservative value derived from the main-pulse.
 
As explained in details in \cite{sw12,sck+13}, the combination of two
pulsars leads to a significant suppression of the long tails in the
probability density function. Assuming that $\hat\xi$ is only weakly
dependent on the pulsar mass, \psrs{} give a combined limit for
strongly self-gravitating bodies of
%---------------------------------------------------------------------
\begin{equation}\label{eq:xi_psr}
  |\hxi| < 3.9 \times 10^{-9} \,, \quad \mbox{(95\% CL)} \,.
\end{equation}
%---------------------------------------------------------------------
The limit (\ref{eq:xi_psr}) is the most constraining one of the three
tests presented in this paper. It is more than three orders of
magnitude better than the limit (\ref{eq:xi_solar}) from the Solar
system and five orders of magnitude better than the limit
(\ref{eq:xi_bnry}) from binary pulsars. This is in accordance with the
$\alpha_2$ and $\hat\alpha_2$ results~\cite{nor87,sw12,sck+13}.

%---------------------------------------------------------------------
%---------------------------------------------------------------------

%=====================================================================
\section{Discussions}
\label{sec:sum}
%=====================================================================

Mach's principle states that the inertial mass of a body is determined
by the total matter distribution in the Universe, so if the matter
distribution is not isotropic, the gravity interaction that a mass
feels can depend on its direction of acceleration~\cite{cs58,cs60}.
The tests presented in this paper are Hughes-Drever-type experiments
which originally were conducted to test a possible anisotropy in mass
through magnetic resonance measurements in
spectroscopy~\cite{hrb60,dre61}.  We note that the constraint on LPI
here is for the gravitational interaction, that is different from the
LPI of Einstein's Equivalence Principle related to special relativity,
see e.g.~\cite{bw02,ahj+07} and the review article \cite{wil06}.

Although we express our limits on the anisotropy of gravity in terms
of the PPN parameter $\xi$ (or its strong-field generalization
$\hxi$), it is quite straightforward to convert them into limits on
the anisotropy of the gravitational constant. From (6.75)
in~\cite{wil93}, one has
%%--------------------------------------------------------------------
\begin{equation}\label{eq:Glocal}
  G_{\rm local} = G_0 \left[ 1 + \xi \left(3+\frac{I}{MR^2}\right) U_{\rm G}
    + \xi \left( {\bf e} \cdot{\bf
      n}_{\rm G} \right)^2\left(1-\frac{3I}{MR^2}\right) U_{\rm G} \right] \,,
\end{equation}
%%--------------------------------------------------------------------
where $G_0$ is the bare gravitational constant; $I$, $M$, and $R$ are
the moment of inertia, mass and radius of a system ${\cal S}$
respectively; ${\bf e}$ is a unit vector pointing from the center of
mass of ${\cal S}$ to the location where $G$ is being measured
(see~\cite{wil93}). The first correction only renormalizes the bare
gravitational constant and is not relevant here. The second correction
contains an anisotropic contribution. For solitary pulsars \psrs{},
they both have $v^2_{\rm G} \sim 5\times10^{-7}$. Hence from
(\ref{eq:xi_psr}), by using $I/MR^2 \simeq 0.4$ for a
typical NS \cite{lp01}, one gets
%------------------------------------------------------------
\begin{equation}\label{eq:deltaG}
  \left| \frac{\Delta G}{G}\right|^{\rm anisotropy} < 4 \times
  10^{-16} \,, \quad \mbox{(95\% CL)}
\end{equation}
%-------------------------------------------------------------------
which is the most constraining limit on the anisotropy of $G$.  It is
four orders of magnitude better than that achievable with LLR in the
foreseeable future.

For any ``quasilinear'' theory of gravity, the PPN parameters satisfy
$\beta = \xi$~\cite{wil73}. Hence for such a theory, a limit on
$\beta$ of ${\cal O}(10^{-9})$ can be drawn, which is six orders of
magnitude more constraining than the limit on $\beta$ from the
anomalous precession of Mercury \cite{wil06}. Nordtvedt developed an
{\it anisotropic PPN framework}~\cite{nor76} and suggested to use the
binary pulsar PSR~B1913+16~\cite{nor75} and LLR~\cite{dbf+94,nor96} to
constrain its parameters. Our result shows that careful profile
analysis of solitary pulsars can constrain some anisotropic PPN
parameters more effectively. The standard model extension of
gravity~\cite{bk06,kt11} has 20 free parameters in the pure-gravity
sector, of which a subset $\bar{s}^{jk}$ appears in a Lagrangian term
similar to (\ref{eq:lagrangianG}) (see (54) in~\cite{bk06}), hence can
be constrained tightly through our tests. We expect a combination of
$\bar{s}^{jk}$ (similar to (97) in~\cite{bk06}) can be constrained to
${\cal O}(10^{-15})$\footnote{See relevant limits from
  LLR~\cite{bcs07} and atom interferometry~\cite{mch+08,cch+09} for
  comparison.}.

At this point we would like to elaborate on the distinction between
the weak-field PPN parameter $\xi$ and its strong-field generalization
$\hxi$.  In GR, $\xi=\hxi=0$, but a distinction is necessary for
alternative gravity theories. Damour and Esposito-Far{\`e}se
explicitly showed that in scalar-tensor theories, the strong
gravitational fields of neutron stars can develop nonperturbative
effects~\cite{de93}. Although scalar-tensor theories have no LPI
violation, one can imagine that similar nonperturbative strong-field
modifications might exist in other theories with LPI violation. If the
strong-field modification is perturbative, one may write an expansion
like,
%%--------------------------------------------------------------------
\begin{equation}
  \hxi = \xi + {\cal K}_1 {\cal C} + {\cal K}_2 {\cal C}^2 + \cdots\,,
\end{equation}
%%--------------------------------------------------------------------
where the compactness ${\cal C}$ (roughly equals the fractional
gravitational binding energy) of a NS (${\cal C}_{\rm NS} \sim 0.2$)
is ${\cal O}(10^5)$ times larger than that of the Sun (${\cal C}_\odot
\sim 10^{-6}$). Hence NSs can probe the coefficients ${\cal K}_i$'s
much more efficiently than the Solar system.

Let us compare the prospects of different tests of LPI in the future.
As mentioned before, the best limit on $\xi$ from superconducting
gravimeter~\cite{wg76} is of ${\cal O}(10^{-3})$. Modern
superconducting gravimeters are more sensitive. They are distributed
around the world, where a total of 25 superconducting gravimeters form
the Global Geodynamics Project (GGP) network~\cite{shi08}.  The
sensitivity of a superconducting gravimeter, installed at a quiet
site, is better than $1\,{\rm nGal} \equiv 10^{-11}\,\mbox{m\,s}^{-2}$
for a one-year measurement, which is less than the seismic noise level
(a few nGal) at the signal frequencies of $\xi$~\cite{shi08}.
However, the test is severely limited by the Earth model and
unremovable Earth noises. Even under optimistic estimations for GGP,
$\xi$ is expected to be constrained to ${\cal O}(10^{-5})$ at best
\cite{shi08}, which is four orders of magnitude away from
(\ref{eq:xi_psr}). The analysis of LLR data usually does not include
the $\xi$ parameter explicitly, but with its analogy with $\alpha_2$,
one can expect a limit of ${\cal O}(10^{-5})$ at
best~\cite{mwt08}. The Solar limit (\ref{eq:xi_solar}) is based on a
long baseline in time (about 4.6\,Gyr), hence it is not going to
improve anymore. In contrast, the limits (\ref{eq:xi_bnry}) and
(\ref{eq:xi_psr}) will continuously improve with $T^{-3/2}$ solely
based on current pulsars, where $T$ is the observational time
span~\cite{sw12,sck+13}.  New telescopes like the Five-hundred-meter
Aperture Spherical Telescope (FAST)~\cite{nlj+11} and the Square
Kilometre Array (SKA)~\cite{sks+09} will provide better sensitivities
in obtaining pulse profiles, that will be very valuable for improving
the limit of $\hxi$ (and also $\hatwo$~\cite{sck+13}), especially for
the weaker pulsar \psrj{}.  In addition, discoveries of new fast
rotating millisecond pulsars through FAST and SKA are expected in
the future, which will enrich our set of testing systems and further
improve the limits.

Let us elaborate on a possible correcting factor to our limits on
$\xi$ and $\hxi$, arising from a more rigorous treatment of the
Galactic mass distribution.  When estimating $U_{\rm G}$, we have
approximated it as $U_{\rm G} \sim v_{\rm G}^2$ which, e.g., at the
location of the Sun gives $U_{\rm G}/c^2 \simeq 5.4\times10^{-7}$.
Mentock pointed out that the dark matter halo might invalidate such an
approximation~\cite{mn96}. However, Gibbons and Will explicitly
showed, by using a Galaxy model with spherically symmetric matter
distribution, that such a correction is roughly a factor of
two~\cite{gw08}.  We use the Galaxy potential model in~\cite{pac90}
that consists of three components, namely the bulge, the disk and the
dark matter halo, and get a factor of 1.86.\footnote{We also tested
  other models for the Galactic mass distribution. They all agree
  within 2\%.}  The results confirm the correcting factor
in~\cite{gw08}, and our limits on $\xi$ and $\hxi$ should be weakened
by this factor (as well as {\it all} previous limits on $\xi$ in
literature).  Nevertheless, the limit (\ref{eq:deltaG}) on the
anisotropy of $G$ will not change because only the product $\xi U_{\rm
  G}$ enters in (\ref{eq:Glocal}).

As a final remark, using the words of \cite{gw08}, also for pulsar
astronomers Whitehead's gravity theory~\cite{whi22} ($\xi = 1$) is
{\it truly dead}.

%=====================================================================
\section*{Acknowledgments}

We thank Nicolas Caballero, David Champion, and Michael Kramer for
valuable discussions. We are grateful to Aris Noutsos for reading the
manuscript. Lijing Shao is supported by China Scholarship Council
(CSC). This research has made use of NASA's Astrophysics Data System.

%=====================================================================

%=====================================================================

\end{document}